\documentclass[showpacs,preprintnumbers,amsmath,amssymb,pra]{revtex4}

\usepackage{graphicx}
\usepackage{dcolumn}
\usepackage{bm}

\begin{document}
\title{Cold Atom Clocks, Precision Oscillators and Fundamental Tests}

\author{S. Bize}
\author{P. Wolf}
    \altaffiliation[Also at: ]{Bureau International des Poids et Mesures, Pavillon de Breteuil, 92312 S\`evres Cedex, France}
\author{M. Abgrall}
\author{L. Cacciapuoti}
\author{A. Clairon}
\author{J. Gr{\"u}nert}
\author{Ph. Laurent}
\author{P. Lemonde}
\author{I. Maksimovic}
\author{C. Mandache}
    \altaffiliation[Also at: ]{Institutul National de Fizica Laserilor, Plasmei si Radiatiei, P.O. Box MG36, Bucaresti, Magurele, Romania}
\author{H. Marion}
\author{F. Pereira Dos Santos}
\author{P. Rosenbusch}
\author{G. Santarelli}
\author{Y. Sortais}
\author{C. Vian}
\author{S. Zhang}
    \affiliation{BNM-SYRTE, Observatoire de Paris, 61 avenue de l'Observatoire, 75014 Paris, France}
\author{C. Salomon}
    \affiliation{Laboratoire Kastler Brossel, 24 rue Lhomond, F-75231 Paris cedex 05, France}
\author{A.N. Luiten}
\author{M.E. Tobar}
    \affiliation{University of Western Australia, School of Physics, Nedlands 6907 WA, Australia}

%
%
%

\begin{abstract}
    We describe two experimental tests of the Equivalence Principle that are
    based on frequency measurements between precision oscillators
    and/or highly accurate atomic frequency standards. Based on
    comparisons between the hyperfine frequencies of $^{87}$Rb and
    $^{133}$Cs in atomic fountains, the first experiment constrains
    the stability of fundamental constants. The second experiment
    is based on a comparison between a cryogenic sapphire
    oscillator and a hydrogen maser. It tests Local Lorentz
    Invariance. In both cases, we report recent results
    which improve significantly over previous experiments.
\end{abstract}

\maketitle              

    \section{Introduction}

    Einstein's Equivalence Principle (EEP) is at the heart of
    special and general relativity and a cornerstone
    of modern physics. The central importance of this postulate in
    modern theory has motivated tremendous work to experimentally
    test EEP \cite{Will}. Additionally, nearly all unification theories (in particular string
    theories) violate EEP at some level \cite{Marciano84,Damour94,Damour02} which further motivates
    experimental searches for such violations. A third motivation comes from a recent
    analysis of absorption systems in the spectra of distant
    quasars \cite{Webb01} which seems to indicate a variation of the fine-structure constant $\alpha$
    over cosmological timescale, in violation of EEP.

    EEP equivalence principle is made of three constituent elements. The Weak
    Equivalence Principle (WEP) postulates that \emph{trajectories of
    neutral freely falling bodies are independent of their structure and
    composition}. Local Lorentz Invariance (LLI) postulates that
    \emph{in any local freely falling reference frame, the result of a non gravitational measurement is independent of the velocity of the
    frame}. Finally, Local Position Invariance (LPI) postulates that
    \emph{in any local freely falling reference frame, the result of a
    non gravitational measurement is independent of where and when it
    is performed}. The experiments described here use precision
    oscillators and atomic clocks to test LLI and LPI.

    In a first section, experiments testing LPI are described.
    In these experiments, frequencies of atomic transitions are
    compared to each other in a local atomic clock comparison. The
    measurements are repeated over a few years. LPI implies that
    these measurements give consistently the same answer, a prediction which is directly tested. These experiments can
    be further interpreted as testing the stability of fundamental
    constants, if one recognizes that any atomic transition frequency can (at least in principle) be expressed as a function of
    properties of elementary particles and parameters of fundamental
    interactions. Such interpretation requires additional input
    from theoretical calculations of atomic frequencies. In this
    first section, after reviewing some of these
    calculations, we describe a comparison between $^{87}$Rb and
    $^{133}$Cs hyperfine frequencies in atomic fountains. We
    give new and significant constraints to the stability of
    fundamental constants based on the results of these
    experiments.

    In a second part, a test of LLI is described.
    The frequency of a macroscopic cryogenic sapphire resonator
    is measured against a hydrogen maser as a function of
    time. We look for sidereal and and semi-sidereal modulations of
    the measured frequency which would indicate a violation of LLI depending on the speed and orientation of the laboratory
    frame with respect to a preferred frame. First the Robertson,
    Mansouri and Sexl theoretical framework is described as
    a basis to interpret the experiments. Then new results improving on previous experiments are reported.


\section{Test of Local Position Invariance. Stability of fundamental constants}

    \subsection{Theory}
    Tests described here are based on highly
    accurate comparisons of atomic energies. In principle at
    least, it is possible to express any atomic energy as a
    function of the elementary particle properties and the
    coupling constants of fundamental interactions using Quantum
    Electro-Dynamics (QED) and Quantum Chromo-Dynamics (QCD). As a
    consequence, it is possible to deduced a constrain to the
    variation of fundamental constants from a measurement of the stability of the ratio between
    various atomic frequencies.

    Different types of atomic transitions are linked
    to different fundamental constants. The hyperfine
    frequency
    in a given electronic state of alkali-like atoms (involved for instance in $^{133}$Cs, $^{87}$Rb \cite{Marion2003},
    $^{199}$Hg$^{+}$ \cite{Prestage95,Berkeland1998}, $^{171}$Yb$^{+}$ microwave clocks) can be approximated by:
    \begin{equation}\label{eq:hyperfine_energy}
    \nu_{\mathrm{hfs}}^{(i)} \simeq R_\infty
    c \times\mathcal{A}_{\mathrm{hfs}}^{(i)} \times
    g^{(i)}\left(\frac{m_e}{m_p}\right)~\alpha^2~F_{\mathrm{hfs}}^{(i)}(\alpha),
    \end{equation}
    where $R_\infty$ is the Rydberg constant, $c$ the speed of
    light, $g^{(i)}$ the nuclear g-factor, $m_e/m_p$ the electron to
    proton mass ratio and $\alpha$ the fine structure constant.
    In this equation, the dimension is given by $R_\infty
    c$, the atomic unit of frequency. $\mathcal{A}_{\mathrm{hfs}}^{(i)}$ is a numerical factor which depends on
    each particular atom. $F_{\mathrm{hfs}}^{(i)}(\alpha)$ is a relativistic correction
    factor to the motion of the valence electron in the vicinity of the nucleus. This factor
    strongly depends on the atomic number $Z$ and has a major
    contribution for heavy nuclei. The superscript $(i)$ indicates that the quantity depends
    on each particular atom. Similarly, the frequency of an
    electronic transition (involved in H \cite{Niering00}, $^{40}$Ca \cite{Helmcke2003}, $^{199}$Hg$^{+}$ \cite{Bize03}, $^{171}$Yb$^{+}$ \cite{Stenger2001,Peik2003} optical clocks) can be approximated by
    \begin{equation}\label{eq:electronic_energy}
    \nu_{\mathrm{elec}}^{(i)} \simeq R_\infty
    c \times \mathcal{A}_{\mathrm{elec}}^{(i)} \times F_{\mathrm{elec}}^{(i)}(\alpha).
    \end{equation}
    Again, the dimension is given by $R_\infty c$.
    $\mathcal{A}_{\mathrm{elec}}^{(i)}$ is a numerical factor.
    $F_{\mathrm{elec}}^{(i)}(\alpha)$ is a function of $\alpha$ which
    accounts for relativistic effects, spin-orbit couplings and many-body effects \footnote{It should be noted that in general the
    energy of an electronic transition has in fact a
    contribution from the hyperfine interaction. However, this
    contribution is a small fraction of the total transition energy and thus carries no significant sensitivity to a variation of fundamental constants.
    The same applies to higher order terms in the expression of the hyperfine energy (eq. \ref{eq:hyperfine_energy}). A precision of 1 to 10 $\%$
    on the sensitivity is sufficient to interpret current experiments.}.
    From equations \ref{eq:hyperfine_energy} and
    \ref{eq:electronic_energy}, it is possible to calculate the
    ratio between the frequencies in atomic species $(i)$
    and $(ii)$, depending on the type of transition involved:
    \begin{eqnarray}
    \label{eq:various_ratios1}
    \frac{\nu_{\mathrm{elec}}^{(ii)}}{\nu_{\mathrm{elec}}^{(i)}} \propto
    \frac{F_{\mathrm{elec}}^{(ii)}(\alpha)}{F_{\mathrm{elec}}^{(i)}(\alpha)}\\
    \label{eq:various_ratios2}
    \frac{\nu_{\mathrm{hfs}}^{(ii)}}{\nu_{\mathrm{elec}}^{(i)}} \propto
    g^{(ii)}\frac{m_e}{m_p}\alpha^2\frac{F_{\mathrm{hfs}}^{(ii)}(\alpha)}{F_{\mathrm{elec}}^{(i)}(\alpha)}\\
    \label{eq:various_ratios3}
    \frac{\nu_{\mathrm{hfs}}^{(ii)}}{\nu_{\mathrm{hfs}}^{(i)}} \propto
    \frac{g^{(ii)}}{g^{(i)}}\frac{F_{\mathrm{hfs}}^{(ii)}(\alpha)}{F_{\mathrm{hfs}}^{(i)}(\alpha)}.
    \end{eqnarray}
    The product of constants $R_\infty c$ cancels out in the ratio
    of two atomic frequencies and only dimensionless factors are left.
    Also, numerical factors that are irrelevant to the present discussion have been
    omitted. Already, the different sensitivity of the various
    type of comparison can be seen in these equations. Comparisons
    between electronic transitions (equation \ref{eq:various_ratios1}) are only sensitive to $\alpha$.
    Comparisons between hyperfine transitions (equation
    \ref{eq:various_ratios3}) are sensitive both to $\alpha$
    and the strong interaction through the nuclear g-factors. Comparisons between an electronic transition and a hyperfine transition (equation
    \ref{eq:various_ratios2}) are sensitive to $\alpha$, to the strong interaction and to the electron
    mass.

    The sensitivity of a given atomic transition to the variation of
    fundamental constants can be derived from equation \ref{eq:hyperfine_energy} and
    \ref{eq:electronic_energy}:
    \begin{equation}\label{eq:sensitivity_hfs}
        \delta \ln \left(\frac{\nu_{\mathrm{hfs}}^{(i)}}{R_{\infty}c}\right)
        \simeq
        \frac{\delta g^{(i)}}{g^{(i)}}+\frac{\delta(m_e/m_p)}{(m_e/m_p)}+\left(2+\alpha \frac{\partial}{\partial\alpha}\ln
        F_{\mathrm{hfs}}^{(i)}(\alpha)\right)
        \times\frac{\delta\alpha}{\alpha}
    \end{equation}
    \begin{equation}\label{eq:sensitivity_ele}
        \delta \ln \left(\frac{\nu_{\mathrm{elec}}^{(i)}}{R_{\infty}c}\right)
        \simeq\left(\alpha \frac{\partial}{\partial\alpha}\ln
        F_{\mathrm{elec}}^{(i)}(\alpha)\right)
        \times\frac{\delta\alpha}{\alpha}
    \end{equation}
    In recent work \cite{Flambaum2003}, it has been suggested that this analysis can
    be pushed one step further by linking the g-factors
    $g^{(i)}$ and the proton mass $m_p$ to fundamental parameters,
    namely the mass scale of QCD $\Lambda_{\mathrm{QCD}}$ and the quark
    mass $m_{q}=(m_{u}+m_{d})/2$. Within this framework, any
    atomic frequency comparison can be interpreted as testing the
    stability of 3 dimensionless fundamental constants: $\alpha$,
    $m_{q}/\Lambda_{\mathrm{QCD}}$ and $m_{e}/\Lambda_{\mathrm{QCD}}$. For any
    transition we can write:
   \begin{equation}\label{eq:sensitivity}
        \delta \ln \left(\frac{\nu^{(i)}}{R_{\infty}c}\right)
        \simeq K^{(i)}_{\alpha}\times\frac{\delta \alpha}{\alpha}+K^{(i)}_{q}\times\frac{\delta (m_{q}/\Lambda_{\mathrm{QCD}})}{(m_{q}/\Lambda_{\mathrm{QCD}})}
        +K^{(i)}_{e}\times\frac{\delta (m_{e}/\Lambda_{\mathrm{QCD}})}{(m_{e}/\Lambda_{\mathrm{QCD}})}
    \end{equation}
    where the superscript $(i)$ again indicates that the
    coefficient depends on each particular transition. Hyperfine transitions are sensitive to all three fundamental
    constants ($K^{(i)}_{\alpha},K^{(i)}_{q}\neq 0$; $K^{(i)}_{e} \simeq 1$).
    For electronic transitions, we have $K^{(i)}_{e},K^{(i)}_{q} \simeq 0$ and therefore
    they are essentially sensitive to variation of $\alpha$. Four well
    chosen atomic transitions constraining the stability of 3 independent
    frequency ratios are enough to constrain independently the
    stability of the three fundamental constants $\alpha$,
    $m_{q}/\Lambda_{\mathrm{QCD}}$ and $m_{e}/\Lambda_{\mathrm{QCD}}$. From these
    equations, it can
    be seen that at least two different hyperfine transitions must
    be involved to independently constrain $m_{q}/\Lambda_{\mathrm{QCD}}$ and
    $m_{e}/\Lambda_{\mathrm{QCD}}$ which emphasizes the need to maintain and
    improve highly accurate microwave atomic clocks \footnote{In principle, it is also possible
    to use a vibrational molecular transition with $K^{(i)}_{\alpha}\simeq 0$ and $K^{(i)}_{e}\simeq
    1/2$.}. With more than 4 well chosen atomic clocks, redundancy
    is achieved which means that a non vanishing variation of fundamental
    constants can be identified by a clear signature.

    Calculations of the coefficients have now been done for a large
    number of atomic species \cite{Prestage95,Flambaum2003,Dzuba99,Dzuba99a,Dzuba2000,Karshenboim00,Dzuba2003}. For hyperfine transitions in
    $^{87}$Rb, $^{133}$Cs and $^{199}$Hg$^{+}$ the most recent calculation
    gives \cite{Flambaum2003}:
    \begin{eqnarray}
    \label{eq:sensitivity_Rb}
        \delta \ln \left(\frac{\nu_{\mathrm{hfs}}(^{87}\mathrm{Rb})}{R_{\infty}c}\right)
        \simeq 2.34\frac{\delta \alpha}{\alpha}-0.064\frac{\delta (m_{q}/\Lambda_{\mathrm{QCD}})}{(m_{q}/\Lambda_{\mathrm{QCD}})}
        +\frac{\delta
        (m_{e}/\Lambda_{\mathrm{QCD}})}{(m_{e}/\Lambda_{\mathrm{QCD}})}\\
   \label{eq:sensitivity_Cs}
        \delta \ln \left(\frac{\nu_{\mathrm{hfs}}(^{133}\mathrm{Cs})}{R_{\infty}c}\right)
        \simeq 2.83\frac{\delta \alpha}{\alpha}+0.11\frac{\delta (m_{q}/\Lambda_{\mathrm{QCD}})}{(m_{q}/\Lambda_{\mathrm{QCD}})}
        +\frac{\delta
        (m_{e}/\Lambda_{\mathrm{QCD}})}{(m_{e}/\Lambda_{\mathrm{QCD}})}\\
    \label{eq:sensitivity_Hg_hfs}
        \delta \ln \left(\frac{\nu_{\mathrm{hfs}}(^{199}\mathrm{Hg}^{+})}{R_{\infty}c}\right)
        \simeq 4.3\frac{\delta \alpha}{\alpha}-0.02\frac{\delta (m_{q}/\Lambda_{\mathrm{QCD}})}{(m_{q}/\Lambda_{\mathrm{QCD}})}
        +\frac{\delta (m_{e}/\Lambda_{\mathrm{QCD}})}{(m_{e}/\Lambda_{\mathrm{QCD}})}
    \end{eqnarray}
    In each case, the most recent and precise value for the $K_{\alpha}^{(i)}$ coefficient given here is in
    good agreement with earlier calculations \cite{Prestage95}. For electronic
    transitions in H, $^{40}$Ca
    and $^{199}$Hg$^{+}$, we have:
       \begin{eqnarray}
        \label{eq:sensitivity_H}
        \delta \ln \left(\frac{\nu_{\mathrm{elec}}(\mathrm{H})}{R_{\infty}c}\right)
        \simeq 0.\\
        \label{eq:sensitivity_Ca}
        \delta \ln \left(\frac{\nu_{\mathrm{elec}}(^{40}\mathrm{Ca})}{R_{\infty}c}\right)
        \simeq 0.03\frac{\delta \alpha}{\alpha}\\
        \label{eq:sensitivity_Hg}
        \delta \ln \left(\frac{\nu_{\mathrm{elec}}(^{199}\mathrm{Hg}^{+})}{R_{\infty}c}\right)
        \simeq -3.2\frac{\delta \alpha}{\alpha}
    \end{eqnarray}
    A new generation of laser cooled optical clocks is now
    under development in several groups ($^{87}$Sr \cite{Courtillot2003,Katori2003,Takamoto2003}, $^{171}$Yb,
    $^{27}$Al$^{+}$ \cite{Wineland2003},...). This work will significantly improve the stringency and redundancy of this test of LPI.

    \subsection{Experiments with $^{87}$Rb and $^{133}$Cs fountain clocks }

\begin{figure}[htb]
\begin{center}
\includegraphics[height=3.5cm]{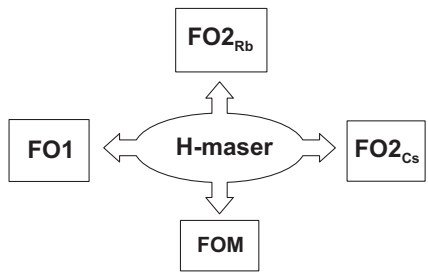}
\end{center}
\caption{BNM-SYRTE clock ensemble.  A single 100\,MHz signal from
a H-maser is used for frequency comparisons and is distributed to
each of the microwave synthesizers of the $^{133}$Cs (FO1, FOM,
FO2$_{\mathrm{Cs}}$) and $^{87}$Rb fountain clocks. In 2001, the
Rb fountain has been upgraded and is now a dual fountain using
alternately rubidium (FO2$_{\mathrm{Rb}}$) or cesium atoms
(FO2$_{\mathrm{Cs}}$). } \label{fig:clocks}
\end{figure}

    In these experiments, three atomic fountains are compared to each
other, using a hydrogen maser (H-maser) as a flywheel oscillator
(Fig.\ref{fig:clocks}). Two fountains, a transportable fountain
FOM, and FO1 \cite{FO1} are using cesium atoms. The third fountain
is a dual fountain (FO2) \cite{Bize01}, operating alternately with
rubidium (FO2$_{\mathrm{Rb}}$) and cesium (FO2$_{\mathrm{Cs}}$).
These fountains have been continuously upgraded in order to
improve their accuracy from $2\times 10^{-15}$ in 1998 to $8\times
10^{-16}$ for cesium and from $1.3 \times 10^{-14}$ \cite{Bize99}
to $6\times 10^{-16}$ for rubidium.

Fountain clocks operate as follows. First, atoms are collected and
laser cooled in an optical molasses or in a magneto-optical trap
in $0.3$ to $0.6$ s. Atoms are launched upwards, and selected in
the clock level ($m_F=0$) by a combination of microwave and laser
pulses. Then, atoms interact twice with a microwave field tuned
near the hyperfine frequency, in a Ramsey interrogation scheme.
The microwave field is synthesized from a low phase noise 100 MHz
signal from a quartz oscillator, which is phase locked to the
reference signal of the H-maser (Fig.\ref{fig:clocks}). After the
microwave interactions, the population of each hyperfine state is
measured using light induced fluorescence. This provides a
measurement of the transition probability as a function of
microwave detuning. Successive measurements are used to steer the
average microwave field to the frequency of the atomic resonance
using a digital servo system. The output of the servo provides a
direct measurement of the frequency difference between the H-maser
and the fountain clock.

The three fountains have different geometries and operating
conditions: the number of detected atoms ranges from $3\times
10^{5}$ to $2\times 10^{6}$ at a temperature of $\sim 1\,\mu$K,
the fountain cycle duration from 1.1 to 1.6 s. The Ramsey
resonance width is between 0.9 and $1.2$ Hz. In measurements
reported here the fractional frequency instability is
$(1-2)\times10^{-13}\tau^{-1/2}$, where $\tau$ is the averaging
time in seconds. Fountain comparisons have a typical resolution of
$\sim 10^{-15}$ for a 12 hour integration, and each of the four
data campaigns lasts from 1 to 2 months during which an accuracy
evaluation of each fountain is performed.

\begin{figure}[htb]
\begin{center}
\includegraphics[height=8cm]{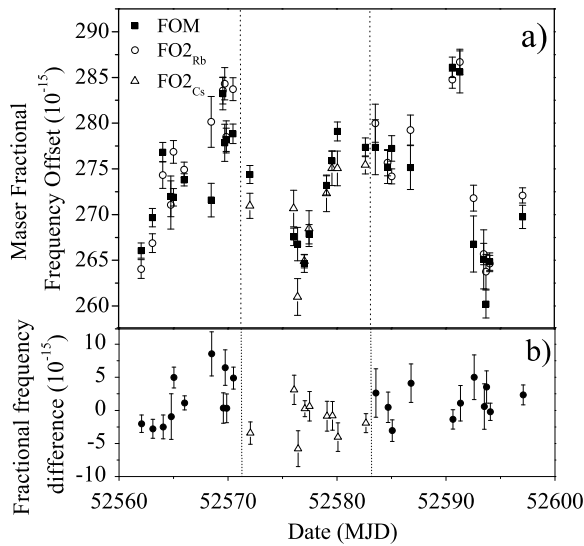}
\end{center}
\caption{The 2002 frequency comparison data. a) H-maser fractional
frequency offset versus FOM (black squares), and alternately
versus FO2$_{\mathrm{Rb}}$ (open circles) and FO2$_{\mathrm{Cs}}$
(open triangles between dotted lines). b) Fractional frequency
differences. Between dotted lines, Cs-Cs comparisons, outside
Rb-Cs comparisons. Error bars are purely statistical. They
correspond to the Allan standard deviation of the comparisons and
do not include contributions from fluctuations of systematic
shifts of Table \ref{tab:Budget}.} \label{fig:compRbCs}
\end{figure}
The 2002 measurements are presented in Fig.\ref{fig:compRbCs},
which displays the maser fractional frequency offset, measured by
the Cs fountains FOM and FO2$_{\mathrm{Cs}}$. Also shown is the
H-maser frequency offset measured by the Rb fountain
FO2$_{\mathrm{Rb}}$ where the Rb hyperfine frequency is
conventionally chosen to be
$\nu_{\mathrm{Rb}}(1999)=6\,834\,682\,610.904\,333\,$Hz, our 1999
value. The data are corrected for the systematic frequency shifts
listed in Table \ref{tab:Budget}. The H-maser frequency exhibits
fractional fluctuations on the order of $10^{-14}$ over a few
days, ten times larger than the typical statistical uncertainty
resulting from the instability of the fountain clocks. In order to
reject the H-maser frequency fluctuations, the fountain data are
recorded simultaneously (within a few minutes). The fractional
frequency differences plotted in Fig.\ref{fig:compRbCs}\,b
illustrate the efficiency of this rejection. FO2 is operated
alternately with Rb and Cs, allowing both Rb-Cs comparisons and
Cs-Cs comparisons (central part of Fig.\ref{fig:compRbCs}) to be
performed.

\begin{table}[htb]
\caption{Accuracy budget of the fountains involved in the 2002
measurements (FO2 et FOM).}
\begin{center}
\begin{tabular}{cccc}
\textbf{Fountain} & \multicolumn{1}{|c}{\textbf{FO2$_{\mathrm{Cs}}$}} & \multicolumn{1}{|c}{\textbf{FO2$_{\mathrm{Rb}}$}} & \multicolumn{1}{|c}{\textbf{FOM}}  \\
\hline\hline Effect & \multicolumn{3}{|c}{ Value \& Uncertainty  (10$^{-16})$} \\
\hline 2$^{nd}$\ order Zeeman &  \multicolumn{1}{|r} {$1773.0\pm5.2$} & \multicolumn{1}{|r} {$3207.0 \pm 4.7$} & \multicolumn{1}{|r} {$385.0 \pm 2.9$}\\
\hline Blackbody Radiation & \multicolumn{1}{|r} {$-173.0\pm 2.3$}& \multicolumn{1}{|r} {$-127.0\pm 2.1$}& \multicolumn{1}{|r} {$-186.0\pm 2.5$}\\
\hline {
\begin{tabular}{r}Cold collisions \\
+ cavity pulling
\end{tabular}
}
 & \multicolumn{1}{|r} {$-95.0\pm 4.6$}& \multicolumn{1}{|r} {$0.0\pm 1.0$}& \multicolumn{1}{|r} {$-24.0\pm 4.8$}\\
\hline others &  \multicolumn{1}{|r} {$0.0\pm 3.0$}& \multicolumn{1}{|r} {$0.0\pm 3.0$}& \multicolumn{1}{|r} {$0.0\pm 3.7$}\\
\hline \hline \textbf{Total uncertainty} & \multicolumn{1}{|r}
{\textbf 8} & \multicolumn{1}{|r} {\textbf 6} &
\multicolumn{1}{|r} {\textbf 8}
\end{tabular}
\end{center}
\label{tab:Budget}
\end{table}
Systematic effects shifting the frequency of the fountain
standards are listed in Table \ref{tab:Budget}. The quantization
magnetic field in the interrogation region is determined with a
$0.1$~nT uncertainty by measuring the frequency of a linear
field-dependent ``Zeeman" transition. The temperature in the
interrogation region is monitored with 5 platinum resistors and
the uncertainty on the black-body radiation frequency shift
corresponds to temperature fluctuations of about 1 K
\cite{Simon98}. Clock frequencies are corrected for the cold
collision and cavity pulling frequency shifts using several
methods \cite{Pereira02,Sortais00}. For Rb, unlike
\cite{Sortais00}, an optical molasses with a small number of atoms
($\sim 5.4\times 10^{6}$) is used. We thus estimate that these two
shifts are smaller than $5\times 10^{-17}$. All other effects do
not contribute significantly and their uncertainties are added
quadratically. We searched for the influence of synchronous
perturbations by changing the timing sequence and the atom launch
height. To search for possible microwave leakage, we changed the
power ($\times9$) in the interrogation microwave cavity. No shift
was found at a resolution of $10^{-15}$. The shift due to residual
coherences and populations in neighboring Zeeman states is
estimated to be less than $10^{-16}$. As shown in \cite{Wolf01},
the shift due to the microwave photon recoil is very similar for
Cs and Rb and smaller than $+ 1.4\times 10^{-16}$. Relativistic
corrections (gravitational redshift and second order Doppler
effect) contribute to less than $10^{-16}$ in the clock
comparisons.

For the Cs-Cs 2002 comparison, we find:
\begin{equation}
\frac{{\nu_{\mathrm{Cs}}^{\mathrm{FO2}}(2002)}-\nu_{\mathrm{Cs}}^{\mathrm{FOM}}(2002)}{\nu_{\mathrm{Cs}}}=+12(6)(12)\times
10^{-16}
\end{equation}
where the first parenthesis reflects the $1\sigma$ statistical
uncertainty, and the second the systematic uncertainty, obtained
by adding quadratically the inaccuracies of the two Cs clocks (see
Table \ref{tab:Budget}). The two Cs fountains are in good
agreement despite their significantly different operating
conditions (see Table \ref{tab:Budget}), showing that systematic
effects are well understood at the $10^{-15}$ level.

In 2002, the $^{87}$Rb frequency measured with respect to the
average $^{133}$Cs frequency is found to be:
\begin{equation}
\nu_{\mathrm{Rb}}(2002)=6\,834\,682\,610.904\,324(4)(7)
\,\mathrm{Hz}
\end{equation}
where the error bars now include FO2$_{\mathrm{Rb}}$,
FO2$_{\mathrm{Cs}}$ and FOM uncertainties. This is the most
accurate frequency measurement to date.

\begin{figure}[htb]
\begin{center}
\includegraphics[height=6cm]{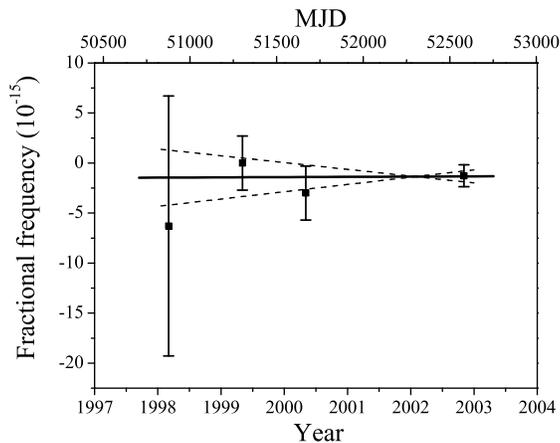}
\end{center}
\caption{Measured $^{87}$Rb frequencies referenced to the
$^{133}$Cs fountains over 57 months. The 1999 measurement value
($\nu_{\mathrm{Rb}}(1999)=6\,834\,682\,610.904\,333\,$Hz) is
conventionally used as reference. A weighted linear fit to the
data gives
$\frac{d}{dt}\ln\left(\frac{\nu_{\mathrm{Rb}}}{\nu_{\mathrm{Cs}}}\right)=(0.2\pm
7.0)\times 10^{-16}\,\mathrm{yr}^{-1}$. Dotted lines correspond to
the $1\sigma$ slope uncertainty.} \label{fig:alphapoint}
\end{figure}
In Fig.\ref{fig:alphapoint} are plotted all our Rb-Cs frequency
comparisons. Except for the less precise 1998 data \cite{Bize99},
two Cs fountains were used together to perform the Rb
measurements. The uncertainties for the 1999 and 2000 measurements
were $2.7\times 10^{-15}$, because of lower clock accuracy and
lack of rigorous simultaneity in the earlier frequency comparisons
\cite{Bize01}. A weighted linear fit to the data in
Fig.\ref{fig:alphapoint} determines how our measurements constrain
a possible time variation of $
\nu_{\mathrm{Rb}}/\nu_{\mathrm{Cs}}$. We find:
\begin{equation} \label{eq:alpha1}
\frac{d}{dt}\ln\left(\frac{\nu_{\mathrm{Rb}}}{\nu_{\mathrm{Cs}}}\right)=(0.2\pm7.0)\times
10^{-16}\,\mathrm{yr}^{-1}
\end{equation}
which represents a 5-fold improvement over our previous results
\cite{Bize01} and a 100-fold improvement over the Hg$^+$-H
hyperfine energy comparison \cite{Prestage95}. Using equation
\ref{eq:sensitivity_hfs} and the sensitivity to $\alpha$ in
equation \ref{eq:sensitivity_Rb} and \ref{eq:sensitivity_Cs}, we
find that this results implies the following constraint:
\begin{equation} \label{eq:alpha2}
\frac{d}{dt}\ln\left(\frac{g_{\mathrm{Cs}}}{g_{\mathrm{Rb}}}\alpha^{0.49}\right)=(0.2\pm7.0)\times
10^{-16}\,\mathrm{yr}^{-1}.
\end{equation}
Using the link between g-factors, $m_{q}$ and
$\Lambda_{\mathrm{QCD}}$ (ref. \cite{Flambaum2003}, eq.
\ref{eq:sensitivity_Rb} and \ref{eq:sensitivity_Cs}), we get:
\begin{equation} \label{eq:alpha3}
\frac{d}{dt}\ln\left(\alpha^{0.49}\left[m_{q}/\Lambda_{\mathrm{QCD}}
 \right]^{0.17}\right)=(0.2\pm7.0)\times 10^{-16}\,\mathrm{yr}^{-1}.
\end{equation}
A comparison between the single mercury $^{199}$Hg$^{+}$ ion
optical clock and the $^{133}$Cs hyperfine splitting has been
recently reported by the NIST group \cite{Bize03} which (according
to eq. \ref{eq:sensitivity_Cs} and \ref{eq:sensitivity_Hg})
constrains the stability of
$\alpha^{6.0}[m_{e}/\Lambda_{\mathrm{QCD}}][m_{q}/\Lambda_{\mathrm{QCD}}]^{0.1}$
at the level of $7\times10^{-15}$~yr$^{-1}$.

\section{Tests of Local Lorentz Invariance}
\index{Lorentz Invariance}

Local Lorentz Invariance is one of the constituent elements of the
Einstein Equivalence Principle (see Sect. 1) and therefore one of
the cornerstones of modern physics. It is the fundamental
hypothesis of special relativity and is related to the "constancy
of the speed of light". The central importance of this postulate
in modern physics has motivated tremendous work to experimentally
test LLI \cite{Will,KostoB}. Additionally, nearly all unification
theories (in particular string theory) violate the EEP at some
level \cite{Damour1} which further motivates experimental searches
for such violations of the universality of free fall
\cite{Damour94} and of Lorentz invariance \cite{Kosto1,Kosto2}.

We report here on experimental tests of LLI using a cryogenic
sapphire oscillator and a hydrogen maser. The relative frequency
of the two clocks is monitored looking for a Lorentz violating
signal which would modulate that frequency at, typically, sidereal
and semi-sidereal periods due to the movement of the lab with the
rotation of the Earth. We set limits on parameters that describe
such Lorentz violating effects, improving our previous results \cite{Wolf}
by a factor 2 and the best other results \cite{Schiller,Muller} by up to a factor 70.

Many modern experiments that test LLI rely essentially on the
stability of atomic clocks and macroscopic resonators
\cite{Brillet,KT,Hils,Schiller,Wolf,Lipa,Muller}, therefore
improvements in oscillator technology have gone hand in hand with
improved tests of LLI. Our experiment is no exception, the large
improvements being a direct result of the excellent stability of
our cryogenic sapphire oscillator. Additionally its operation at a
microwave frequency allows a direct comparison to a hydrogen maser
which provides a highly stable and reliable reference frequency.

Numerous test theories that allow the modeling and interpretation
of experiments that test LLI have been developed. Kinematical
frameworks \cite{Robertson,MaS} postulate a simple parametrisation
of the Lorentz transformations with experiments setting limits on
the deviation of those parameters from their special relativistic
values. A more fundamental approach is offered by theories that
parametrise the coupling between gravitational and
non-gravitational fields (TH$\epsilon\mu$
\cite{LightLee,Will,Blanchet} or $\chi$g \cite{Ni} formalisms)
which allow the comparison of experiments that test different
aspects of the EEP. Finally, formalisms based on string theory
\cite{KostoB,Damour1,Damour94,Kosto1} have the advantage of being
well motivated by theories of physics that are at present the only
candidates for a unification of gravity and the other fundamental
forces of nature.

\subsection{Theory}

Owing to their simplicity the kinematical frameworks of
\cite{Robertson,MaS} have been widely used to model and interpret
many previous experiments testing LLI
\cite{Brillet,Hils,Schiller,Wolf,Riis,WP} and we will follow that
route. An analysis based on the more fundamental "Standard Model
Extension" (SME) \cite{Kosto1,KM} is under way and will be
published shortly.

By construction kinematical frameworks do not allow for any
dynamical effects on the measurement apparatus. This implies that
in all inertial frames two clocks of different nature (e.g. based
on different atomic species) run at the same relative rate and two
length standards made of different materials keep their relative
lengths. Coordinates are defined by the clocks and length
standards, and only the transformations between those coordinate
systems are modified. In general this leads to observable effects
on light propagation in moving frames but, by definition, to no
observable effects on clocks and length standards. In particular,
no attempt is made at explaining the underlying physics (e.g.
modified Maxwell and/or Dirac equations) that could lead to
Lorentz violating light propagation but leave e.g. atomic energy
levels unchanged. On the other hand dynamical frameworks (e.g. the
TH$\epsilon\mu$ formalism or the SME) in general use a modified
general Lagrangian that leads to modified Maxwell and Dirac
equations and hence to Lorentz violating light propagation and
atomic properties, which is why they are considered more
fundamental and more complete than the kinematical frameworks.
Furthermore, as shown in \cite{KM}, the SME is kept sufficiently
general to, in fact, encompass the kinematical frameworks and some
other dynamical frameworks (in particular the TH$\epsilon\mu$
formalism) as special cases, although there are no simple and
direct relationships between the respective parameters.

Concerning our experiment the SME allows the calculation of
Lorentz violating effects on the fields inside the sapphire
resonator, on the properties of the sapphire crystal itself and on
the hydrogen maser transition. As shown in \cite{Muller2} the
effect on the sapphire crystal amounts to only a few percent of
the direct effect on the fields, and \cite{KL} show that the
hydrogen $m_F=0 \to m'_F=0$ clock transition is not affected to
first order. Hence the total effect is dominated by the Lorentz
violating properties of the electromagnetic fields inside the
resonator which can be calculated following the principles laid
down in \cite{KM}. An analysis of our experiment in that framework
is currently being carried out, and the results will be published
elsewhere in the near future. In this paper we concentrate on the
analysis using the kinematical framework of Mansouri and Sexl
\cite{MaS}.

\subsection*{The Robertson, Mansouri \& Sexl Framework}

Kinematical frameworks for the description of Lorentz violation
have been pioneered by Robertson \cite{Robertson} and further
refined by Mansouri and Sexl \cite{MaS} and others. Fundamentally
the different versions of these frameworks are equivalent, and
relations between their parameters are readily obtained. As
mentioned above these frameworks postulate generalized
transformations between a preferred frame candidate $\Sigma(T,{\bf
X})$ and a moving frame ${\rm S}(t,{\bf x})$ where it is assumed
that in both frames coordinates are realized by identical
standards. We start from the transformations of \cite{MaS} (in
differential form) for the case where the velocity of $S$ as
measured in $\Sigma$ is along the positive X-axis, and assuming
Einstein synchronization in $S$ (we will be concerned with signal
travel times around closed loops so the choice of synchronization
convention can play no role):

\begin{equation}
dT = {1\over a}\left(dt+{vdx\over c^2}\right); dX = {dx\over
b}+{v\over a}\left(dt+{vdx\over c^2}\right); dY = {dy\over d}; dZ
= {dz\over d} \label{MStransf}
\end{equation}
with $c$ the velocity of light in vacuum in $\Sigma$. Using the
usual expansion of the three parameters $(a \approx
1+\alpha_{\mathrm{MS}}{v^2/c^2} + {\cal O}(4); b \approx
1+\beta_{\mathrm{MS}}{v^2/c^2} + {\cal O}(4); d \approx
1+\delta_{\mathrm{MS}}{v^2/c^2} + {\cal O}(4))$, setting
$c^2dT^2=dX^2+dY^2+dZ^2$ in $\Sigma$, and transforming according
to (\ref{MStransf}) we find the coordinate travel time of a light
signal in S:

\begin{equation}
dt={dl\over c}\left(1-\left(\beta_{\mathrm{MS}}
-\alpha_{\mathrm{MS}} -1 \right){v^2\over c^2} - \left({1\over
2}-\beta_{\mathrm{MS}} +\delta_{\mathrm{MS}} \right){\rm
sin}^2\theta{v^2\over c^2}\right)+{\cal O}(4) \label{MSc}
\end{equation}
where $dl = \sqrt{dx^2+dy^2+dz^2}$ and $\theta$ is the angle
between the direction of light propagation and the velocity {\bf
v} of S in $\Sigma$. In special relativity $\alpha_{\mathrm{MS}} =
-1/2; \beta_{\mathrm{MS}} = 1/2; \delta_{\mathrm{MS}} = 0$ and
(\ref{MStransf}) reduces to the usual Lorentz transformations.
Generally, the best candidate for $\Sigma$ is taken to be the
frame of the cosmic microwave background (CMB) \cite{Fixsen,Lubin}
with the velocity of the solar system in that frame taken as
$v_\odot \approx 377$ km/s, decl. $\approx -6.4 ^\circ $, $RA
\approx 11.2$h.

Michelson-Morley type experiments \cite{MM,Brillet} determine the
coefficient $P_{MM} = (1/2-\beta_{\mathrm{MS}}
+\delta_{\mathrm{MS}})$ of the direction dependent term. For many
years the most stringent limit on that parameter was $|P_{MM}|
\leq 5 \times 10^{-9}$ determined over 23 years ago in an
outstanding experiment \cite{Brillet}. Our experiment confirms
that result with roughly equivalent uncertainty $(2.2 \times
10^{-9})$. Recently an improvement to $|P_{MM}| \leq 1.5 \times
10^{-9}$ has been reported \cite{Muller}. Kennedy-Thorndike
experiments \cite{KT,Hils,Schiller} measure the coefficient
$P_{KT} = (\beta_{\mathrm{MS}} -\alpha_{\mathrm{MS}} -1)$ of the
velocity dependent term. The most stringent limit \cite{Schiller}
on $|P_{KT}|$ has been recently improved from \cite{Hils} by a
factor 3 to $|P_{KT}| \leq 2.1 \times 10^{-5}$. We improve this
result by a factor of 70 to $|P_{KT}| \leq 3.0 \times 10^{-7}$.
Finally clock comparison and Doppler experiments
\cite{Riis,Grieser,WP} measure $\alpha_{\mathrm{MS}}$, currently
limiting it to $|\alpha_{\mathrm{MS}} + 1/2| \leq 8 \times
10^{-7}$. The three types of experiments taken together then
completely characterize any deviation from Lorentz invariance in
this particular test theory, with present limits summarized in
Table \ref{MStab}.

\begin{table}
\caption{Present limits on Lorentz violating parameters in the
framework of \cite{MaS}}
\begin{center}
\renewcommand{\arraystretch}{1.4}
\setlength\tabcolsep{5pt}
\begin{tabular}{cccc}
\hline\noalign{\smallskip}
Reference & $\alpha_{\mathrm{MS}} + 1/2$ & $1/2-\beta_{\mathrm{MS}} +\delta_{\mathrm{MS}}$ & $\beta_{\mathrm{MS}} -\alpha_{\mathrm{MS}} -1$ \\
\noalign{\smallskip} \hline \noalign{\smallskip}
\cite{Riis,Grieser,WP} & $\leq 8 \times 10^{-7}$ & - & -  \\
\cite{Brillet} & - & $\leq 5 \times 10^{-9}$ & - \\
\cite{Muller} & - & $(2.2 \pm 1.5)\times 10^{-9}$ & - \\
\cite{Schiller} & - & - & $(1.9 \pm 2.1)\times 10^{-5}$ \\
our previous results \cite{Wolf} & - & $(1.5 \pm 4.2)\times 10^{-9}$ & $(-3.1 \pm 6.9)\times 10^{-7}$ \\
this work & - & $(1.2 \pm 2.2)\times 10^{-9}$ & $(1.6 \pm 3.0)\times 10^{-7}$ \\
\hline
\end{tabular}
\end{center}
\label{MStab}
\end{table}

\begin{figure}[b]
\begin{center}
\includegraphics[width=9cm]{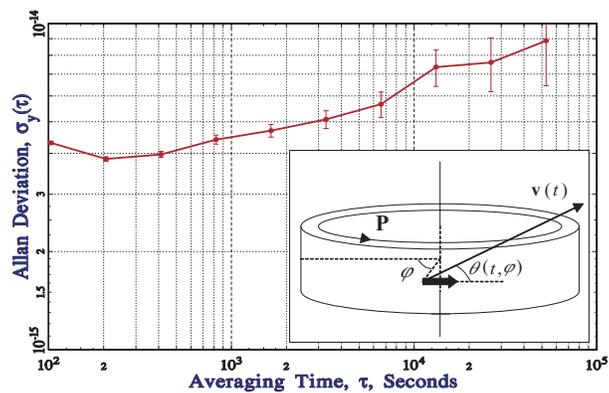}
\end{center}
\caption[]{Typical relative frequency stability of the CSO -
H-maser difference after removal of a linear frequency drift. The
inset is a schematic drawing of the cylindrical sapphire
oscillator with the Poynting vector $\bf P$ in the whispering
gallery (WG) mode, the velocity ${\bf v}(t)$ of the cylinder with
respect to the CMB, and the relevant angles for a photon in the WG
mode.} \label{MMKTfig1}
\end{figure}

Our cryogenic oscillator consists of a sapphire crystal of
cylindrical shape operating in a whispering gallery mode (see Fig.
\ref{MMKTfig1} for a schematic drawing and \cite{Chang,Mann} for a
detailed description). Its coordinate frequency can be expressed
by  $\nu = m/t_c$ where $t_c$ is the coordinate travel time of a
light signal around the circumference of the cylinder (of radius
$r$) and $m$ is a constant. From (\ref{MSc}) the relative
frequency difference between the sapphire oscillator and the
hydrogen maser (which, by definition, realizes coordinate time in
S \cite{masercom}) is

\begin{equation}
{\Delta\nu (t) \over \nu_0} = P_{KT}{v(t)^2\over c^2} +
P_{MM}{v(t)^2\over c^2}{1 \over 2\pi}\int_0^{2 \pi}{\rm
sin}^2\theta (t,\varphi ) d\varphi +{\cal O}(3) \label{MSdff}
\end{equation}
where $\nu_0 = m/(2\pi r/c)$, $v(t)$ is the (time dependent) speed
of the lab in $\Sigma$, and $\varphi$ is the azimuthal angle of
the light signal in the plane of the cylinder. The periodic time
dependence of $v$ and $\theta$ due to the rotation and orbital
motion of the Earth with respect to the CMB frame allow us to set
limits on the two parameters in (\ref{MSdff}) by adjusting the
periodic terms of appropriate frequency and phase (see \cite{Mike}
for calculations of similar effects for several types of
oscillator modes). Given the limited durations of our data sets
($\leq$ 15 days) the dominant periodic terms arise from the
Earth's rotation, so retaining only those we have ${\bf v}(t) =
{\bf u}+{\bf \Omega} \times {\bf R}$ with ${\bf u}$ the velocity
of the solar system with respect to the CMB, ${\bf \Omega}$ the
angular velocity of the Earth, and ${\bf R}$ the geocentric
position of the lab. We then find after some calculation.

\begin{equation}
\begin{array}{cl}
\Delta\nu / \nu_0 &= P_{KT}(H{\rm sin}\lambda )\\
\ & + P_{MM}(A{\rm cos}\lambda + B{\rm cos}(2\lambda)+C{\rm
sin}\lambda+D{\rm sin}\lambda{\rm cos}\lambda+E{\rm
sin}\lambda{\rm cos}(2\lambda))
\end{array}
\label{MSdff2}
\end{equation}
where $\lambda =\Omega t + \phi$, and A-E and $\phi$ are constants
depending on the latitude and longitude of the lab $(\approx 48.7
^\circ$N and $2.33 ^\circ$E for Paris). Numerically $H \approx
-2.6 \times 10^{-9}$, $A \approx -8.8 \times 10^{-8}$, $B \approx
1.8 \times 10^{-7}$, C-E of order $10^{-9}$. We note that in
(\ref{MSdff2}) the dominant time variations of the two
combinations of parameters are in quadrature and at twice the
frequency which indicates that they should decorelate well in the
data analysis allowing a simultaneous determination of the two (as
confirmed by the correlation coefficients given in Sect. 3.2).
Adjusting this simplified model to our data we obtain results that
differ by less than 10\% from the results presented in Sect. 3.2
that were obtained using the complete model ((\ref{MSdff})
including the orbital motion of the Earth).

\subsection{Experimental Results}

The cryogenic sapphire oscillator (CSO) is compared to a
commercial (Datum Inc.) active hydrogen maser whose frequency is
also regularly compared to caesium and rubidium atomic fountain
clocks in the laboratory \cite{Bize01}. The CSO resonant frequency
at 11.932 GHz is compared to the 100 MHz output of the hydrogen
maser. The maser signal is multiplied up to 12 GHz of which the
CSO signal is subtracted. The  remaining $\approx$ 67 MHz signal
is mixed to a synthesizer signal at the same frequency and the low
frequency beat at $\approx$ 64 Hz is counted, giving access to the
frequency difference between the maser and the CSO. The
instability of the comparison chain has been measured and does not
exceed a few parts in $10^{16}$. The typical stability of the
measured CSO - maser frequency after removal of a linear frequency
drift is shown in Fig. \ref{MMKTfig1}. Since September 2002 we are
taking continuous temperature measurements on top of the CSO dewar
and behind the electronics rack. Starting January 2003 we have
implemented an active temperature control of the CSO room and
changed some of the electronics. As a result the diurnal and
semi-diurnal temperature variations during measurement runs
($\approx$ 2 weeks) were greatly reduced to less than
$0.025^\circ$ C in amplitude (best case), and longer and more
reliable data sets became available.

Our previously published results \cite{Wolf} are based on data
sets taken between Nov. 2001 and Sep. 2002 which did not all
include regular temperature monitoring and control. Here we use
only data that was permanently temperature controlled, 13 data
sets in total spanning Sept. 2002 to Aug. 2003, of differing
lengths (5 to 16 days, 140 days in total). The sampling time for
all data sets was $100$ s except two data sets with $\tau_0 = 12$
s. To make the data more manageable we first average all points to
$\tau_0 = 2500$ s. For the data analysis we simultaneously adjust
an offset and a rate (natural frequency drift, typically $\approx
1.7 \times 10^{-18}$ s$^{-1}$) per data set and the two parameters
of the model (\ref{MSdff}). In the model (\ref{MSdff}) we take
into account the rotation of the Earth and the Earth's orbital
motion, the latter contributing little as any constant or linear
terms over the durations of the individual data sets are absorbed
by the adjusted offsets and rates.

When carrying out an ordinary least squares (OLS) adjustment we
note that the residuals have a significantly non-white behavior as
one would expect from the slope of the Allan deviation of Fig.
\ref{MMKTfig1}. The power spectral density (PSD) of the residuals
when fitted with a power law model of the form $S_y(f)=kf^\mu$
yields typically $\mu \approx -1.5$ to $-2$. In the presence of
non-white noise OLS is not the optimal regression method
\cite{lss,Draper} as it can lead to significant underestimation of
the parameter uncertainties \cite{lss}.

\begin{figure}[b]
\begin{center}
\includegraphics[width=9cm]{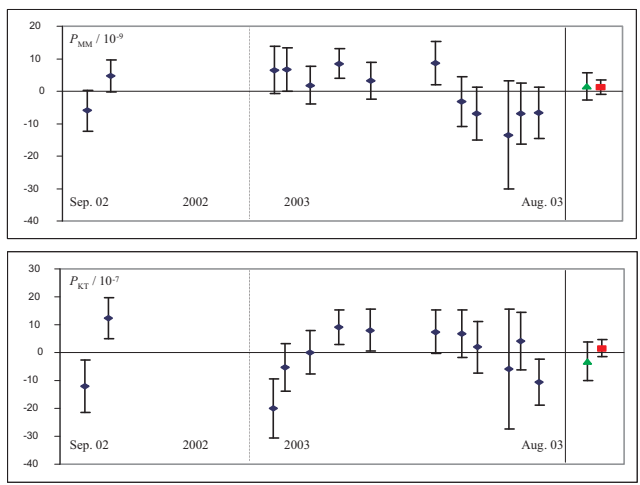}
\end{center}
\caption[]{Values of the two parameters ($P_{KT}$ and $P_{MM}$)
from a fit to each individual data set (blue diamonds) and a
global fit to all the data (red squares). For comparison our
previously published results \cite{Wolf} are also shown (green
triangles). The error bars indicate the combined uncertainties
from statistics and systematic effects.} \label{MMKTfig2}
\end{figure}

An alternative method is weighted least squares (WLS)
\cite{Draper} which allows one to account for non-white noise
processes in the original data by pre-multiplying both sides of
the design equation (our equation (\ref{MSdff}) plus the offsets
and rates) by a weighting matrix containing off diagonal elements.
To determine these off diagonal terms we first carry out OLS and
adjust the $S_y(f)=kf^\mu$ power law model to the PSD of the
post-fit residuals determining a value of $\mu$ for each data set.
We then use these $\mu$ values to construct a weighting matrix
following the method of fractional differencing described, for
example, in \cite{lss}. Figure \ref{MMKTfig2} shows the resulting
values of the two parameters ($P_{KT}$ and $P_{MM}$) for each
individual data set. A global WLS fit of the two parameters and
the 13 offsets and drifts yields $P_{MM} = (1.2\pm 1.9) \times
10^{-9}$ and $P_{KT} = (1.6\pm 2.3) \times 10^{-7}$ ($1\sigma$
uncertainties), with the correlation coefficient between the two
parameters less than 0.01 and all other correlation coefficients
$< 0.06$. The distribution of the 13 individual values around the
ones obtained from the global fit is well compatible with a normal
distribution ($\chi^2$ = 10.7 and 14.6 for $P_{MM}$ and $P_{KT}$
respectively).

Systematic effects at diurnal or semi-diurnal frequencies with the
appropriate phase could mask a putative sidereal signal. The
statistical uncertainties of $P_{MM}$ and $P_{KT}$ obtained from
the WLS fit above correspond to sidereal and semi-sidereal terms
(from (\ref{MSdff2})) of $\approx 7 \times 10^{-16}$ and $\approx
4 \times 10^{-16}$ respectively so any systematic effects
exceeding these limits need to be taken into account in the final
uncertainty. We expect the main contributions to such effects to
arise from temperature, pressure and magnetic field variations
that would affect the hydrogen maser, the CSO and the associated
electronics, and from tilt variations of the CSO which are known
to affect its frequency. Measurements of the tilt variations of
the CSO show amplitudes of 4.6 $\mu$rad and 1.6 $\mu$rad at
diurnal and semi-diurnal frequencies.

To estimate the tilt sensitivity we have intentionally tilted the
oscillator by $\approx$ 5 mrad off its average position which led
to relative frequency variations of $\approx 3 \times 10^{-13}$
from which we deduce a tilt sensitivity of $\approx 6 \times
10^{-17} \mu$rad$^{-1}$. This value corresponds to a worst case
scenario as we expect a quadratic rather than linear frequency
variation for small tilts around the vertical. Even with this
pessimistic estimate diurnal and semi-diurnal frequency variations
due to tilt do not exceed $3 \times 10^{-16}$ and $1 \times
10^{-16}$ respectively and are therefore negligible with respect
to the statistical uncertainties.

In December 2002 we implemented an active temperature
stabilization inside an isolated volume ($\approx 15 {\rm m}^3$)
that included the CSO and all the associated electronics. The
temperature was measured continously in two fixed locations
(behind the electronics rack and on top of the dewar). For the
best data sets the measured temperature variations did not exceed
0.02/0.01 $^\circ$C in amplitude for the diurnal and semi-diurnal
components. In the worst cases (the two 2002 data sets and some
data sets taken during a partial air conditioning failure) the
measured temperature variations could reach 0.26/0.08 $^\circ$C.
When intentionally heating and cooling the CSO lab by $\approx
3^\circ$C we see frequency variations of $\approx 5 \times
10^{-15}$ per $^\circ$C. This is also confirmed when we induce a
large sinusoidal temperature variation ($\approx 1.5 ^\circ$C
amplitude). Using this we can calculate a value for temperature
induced frequency variations at diurnal and semi-diurnal
frequencies for each data set, obtaining values that range from
$\approx 1.3 \times 10^{-15}$ to $\approx 5 \times 10^{-17}$.

The hydrogen maser is kept in a dedicated, environmentally
controlled clock room. Measurements of magnetic field, temperature
and atmospheric pressure in that room and the maser sensitivities
as specified by the manufacturer allow us to exclude any
systematic effects on the maser frequency that would exceed the
statistical uncertainties above and the systematic uncertainties
from temperature variations in the CSO lab.

Our final uncertainties (the error bars in Fig. \ref{MMKTfig2})
are the quadratic sums of the statistical uncertainties from the
WLS adjustment for each data set and the systematic uncertainties
calculated for each data set from (\ref{MSdff2}) and the measured
temperature variations. For the global adjustment we average the
systematic uncertainties from the individual data sets obtaining
$\pm 1.2 \times 10^{-9}$ on $P_{MM}$ and $\pm 1.9 \times 10^{-7}$
on $P_{KT}$. Adding these quadratically to the WLS statistical
uncertainties of the global adjustment we obtain as our final
result $P_{MM} = (1.2\pm 2.2) \times 10^{-9}$ and $P_{KT} =
(1.6\pm 3.0) \times 10^{-7}$ ($1\sigma$ uncertainties).

\section{Conclusion and Outlook}

We have reported on two different tests of the Einstein
Equivalence Principle (EEP) using the comparison of atomic clocks
with different atomic species on one hand, and the comparison of
an atomic clock and a cryogenic sapphire cavity oscillator on the
other. The two experiments are interpreted as testing Local
Position Invariance (LPI) and Local Lorentz Invariance (LLI)
respectively which are both constituent elements of the EEP.

The test of LPI is based on the comparison of the hyperfine
transitions in $^{87}$Rb and $^{133}$Cs using atomic fountains
that presently reach uncertainties of $(6 - 8) \times 10^{-16}$.
Such measurements were repeated over the last 5 years to search for
 a time variation that would indicate a violation of LPI.
 Our present results limit a linear variation to
 $\frac{d}{dt}\ln\left(\frac{\nu_{\mathrm{Rb}}}{\nu_{\mathrm{Cs}}}\right)=(0.2\pm7.0)\times 10^{-16}\,\mathrm{yr}^{-1}$
 which represents a 5-fold improvement over our previous results \cite{Bize01} and a 100-fold
improvement over the Hg$^+$-H hyperfine energy comparison
\cite{Prestage95}. When interpreting the results as a limit on the
time variation of fundamental constants (c.f. Sect. 2.1) we obtain
\begin{equation}
\left|0.49
\frac{\dot{\alpha}}{\alpha}+0.17\frac{\dot{m_{q\Lambda}}}{m_{q\Lambda}}\right|
\leq 7 \times 10^{-16}\,\mathrm{yr}^{-1}
\end{equation}
where $m_{q\Lambda}$ stands for $m_q/\Lambda_{\mathrm{QCD}}$. By
itself this experiment limits the time variation of a combination
of two of the three fundamental constants of Sect. 2.1. The
$^{199}$Hg$^+$ to $^{133}$Cs comparisons by the NIST group
\cite{Bize03} provide
\begin{equation}
\left|6.0\frac{\dot{\alpha}}{\alpha}+0.1\frac{\dot{m_{q\Lambda}}}{m_{q\Lambda}}+\frac{\dot{m_{e\Lambda}}}{m_{e\Lambda}}\right|
\leq 7 \times 10^{-15}\,\mathrm{yr}^{-1}
\end{equation}
where $m_{e\Lambda}$ stands for $m_e/\Lambda_{\mathrm{QCD}}$.
Combining these two results we have two constraints on the
variation of the three fundamental constants, with the missing
third constraint requiring the comparison over time with a fourth
atomic transition (c.f. Sect. 2.1).

The test of LLI is based on the comparison of a hydrogen maser
clock to a cryogenic sapphire microwave cavity. This experiment
simultaneously constrains two combinations of the three parameters
of the Mansouri and Sexl test theory (previously measured
individually by Michelson-Morley and Kennedy-Thorndike
experiments). We obtain $\delta_{\mathrm{MS}} -
\beta_{\mathrm{MS}} + 1/2 = 1.2(1.9)(1.2) \times 10^{-9}$ which is
of the same order as the best previous results
\cite{Brillet,Muller}, and $\beta_{\mathrm{MS}} -
\alpha_{\mathrm{MS}} - 1 = 1.6(2.3)(1.9)\times 10^{-7}$ which
improves the best previous limit \cite{Schiller} by a factor of 70
(the first bracket indicates the $1\sigma$ uncertainty from
statistics the second from systematic effects). We improve our own
previous results \cite{Wolf} by about a factor 2 due to more and
longer data sets and to improved temperature control of the
experiment. We note that our value on $\delta_{\mathrm{MS}} -
\beta_{\mathrm{MS}} + 1/2$ is compatible with the slightly
significant recent result of \cite{Muller} who obtained
$\delta_{\mathrm{MS}} - \beta_{\mathrm{MS}} + 1/2 = (2.2 \pm 1.5)
\times 10^{-9}$.

As a result of our experiment the Lorentz transformations are
confirmed in this particular test theory with an overall
uncertainty of $\leq 8 \times 10^{-7}$ limited now by the
determination of $\alpha_{\mathrm{MS}}$ from Doppler and clock
comparison experiments \cite{Riis,WP}. This is likely to be
improved in the coming years by experiments such as ACES (Atomic
Clock Ensemble in Space \cite{ACES}) that will compare ground
clocks to clocks on the international space station aiming at a 10
fold improvement on the determination of $\alpha_{\mathrm{MS}}$.

In the future, the two tests of LPI and LLI presented here are
expected to further improve due to improvements in the accuracies
of the atomic clocks involved and due to new experimental
strategies, ultimately leading to space-borne versions of the
experiments.

Ongoing efforts are expected to improve the accuracy of both
$^{87}$Rb and $^{133}$Cs to the $10^{-16}$ level. The
corresponding limit to variations of fundamental constants will
then be decreased to $\sim 10^{-16}$~yr$^{-1}$ or less. Recent
advances in the field of optical frequency metrology will probably
lead optical frequency standards to surpass microwave clocks.
Comparing such standards to each other will provide very stringent
limits to the variation of the fine structure constant $\alpha$.
To keep the constraints to the variation of
$m_{q}/\Lambda_{\mathrm{QCD}}$ and $m_{e}/\Lambda_{\mathrm{QCD}}$
at the same level, further efforts and new methods will have to be
invented to improve microwave clocks. These tests will also
greatly benefit from a new generation of time/frequency transfer
at the $10^{-16}$ level which is currently under development for
the ESA space mission ACES which will fly ultra-stable clocks on
board the international space station in 2006 \cite{ACES} and a
similar project conducted by NASA. These missions will allow
highly precise comparisons between clocks developed in distant
laboratories and based on different atomic species and/or
different technologies.

Concerning the test of LLI, new proposals have been made to use
two orthogonal resonators or two orthogonal modes in the same
sapphire resonator placed on a rotating platform \cite{Mike}. Such
a set-up is likely to improve the tests of LLI by several orders
of magnitude as the relevant time variations will now be at the
rotation frequency ($\approx 0.01 - 0.1$ Hz) which is the range in
which such resonators are the most stable ($\approx$ 100 fold
better stability). Additionally many systematic effects are likely
to cancel between the two orthogonal oscillators or modes and the
remaining ones are likely to be less coupled to the rotation
frequency than to the sidereal frequencies used in our experiment.
Ultimately, it has been proposed \cite{Lammerzahl2001} to conduct
these tests on board an Earth orbiting satellite, again with a
potential gain of several orders of magnitudes over current
limits.


%


\begin{thebibliography}{8.}
\bibitem{masercom} We assume here that local position invariance (c.f. Sect. 2) is sufficiently verified so that the variation of the maser frequency due to the diurnal variation of the local gravitational potential is negligible. Indeed the results of \c
ite{Bauch} imply that such variations should not exceed 2 parts in $10^{-17}$ which is significantly below our noise level.
\end{thebibliography}
\end{document}